\documentclass[aps,prl,reprint,amssymb,amsmath,superscriptaddress]{revtex4-2}
\usepackage{graphicx}
\usepackage{dcolumn}
\usepackage{bm}
\usepackage{subfigure}

\newcommand{\fr}[2]{\frac{\displaystyle #1}{\displaystyle #2}}
\newcommand{\df}[2]{\frac{\displaystyle d#1}{\displaystyle d#2}}

\begin{document}

\title{Bumps, chimera states, and Turing patterns in systems of coupled active rotators}%

\author{Igor Franovi\'c}
\email{franovic@ipb.ac.rs}
\affiliation{Scientific Computing Laboratory, Center for the Study of Complex Systems,
Institute of Physics Belgrade, University of Belgrade, Pregrevica 118, 11080 Belgrade, Serbia}

\author{Oleh E. Omel'chenko}
\email{omelchenko@uni-potsdam.de}
\affiliation{University of Potsdam, Institute of Physics and Astronomy,
Karl-Liebknecht-Str. 24/25, 14476 Potsdam-Golm, Germany}

\author{Matthias Wolfrum}
\email{wolfrum@wias-berlin.de}
\affiliation{Weierstrass Institute, Mohrenstrasse 39, 10117 Berlin, Germany}

\date{\today}

\begin{abstract}
Self-organized coherence-incoherence patterns, called {\em chimera states}, have first been
reported in systems of Kuramoto oscillators. For coupled excitable units, similar patterns
where coherent units are at rest, are called {\em bump states}. Here, we study bumps in an
array of active rotators coupled by non-local attraction and global repulsion. We demonstrate
how they can emerge in a \emph{supercritical} scenario from completely coherent Turing patterns:
a single incoherent unit appears in a homoclinic bifurcation, undergoing subsequent transitions
to quasiperiodic and chaotic behavior, which eventually transforms into extensive chaos with
many incoherent units. We present different types of transitions and explain the formation of coherence-incoherence patterns according to the classical paradigm of short-range activation
and long-range inhibition.
\end{abstract}


\maketitle

Since their discovery in 2002 by Kuramoto \& Battogtokh \cite{KB02} chimera states have attracted remarkable attention. They represent a new type of self-organization phenomenon where identical units in a system with symmetric couplings develop a stable pattern with regions of qualitatively different behavior.
In their original work, Kuramoto and Battogtokh found such patterns with self-organized domains of synchronized (coherent) and non-synchronized (incoherent) oscillators in systems of phase oscillators in a one-dimensional array with nonlocal coupling. After the term
{\em chimera state} was coined by Abrams \& Strogatz \cite{AS04}, it has been used for similar phenomena in a large variety of theoretical models \cite{PJAS21,PA15} and has also been demonstrated in various experiments \cite{TRTSE18,HMRHOS12,LRCS19,ARMS16}. In spite of the abundance of examples of chimera states \cite{OK19,AMSW08,MBP16,MRJZ19,Z20,OOHS13,SZAS16,HKK19}, sometimes even only loosely related to the original phenomenon from phase oscillator systems, an understanding of a principal mechanism leading to their formation is still missing \cite{ZM21}. In \cite{AS04}, it has been pointed out as an intriguing property of chimera states that they {\em "cannot be ascribed to a supercritical instability"} since they always stably coexist with the uniform locked state. Since then, no supercritical scenario leading to the emergence of chimera states has been presented. Some recent progress has been made only for the case of Stuart-Landau oscillators with global nonlinear coupling, where clustering has been identified as a prerequisite for chimera states in such systems \cite{SK15}.

\begin{figure}[ht]
\centering
\includegraphics[width=0.9\columnwidth]{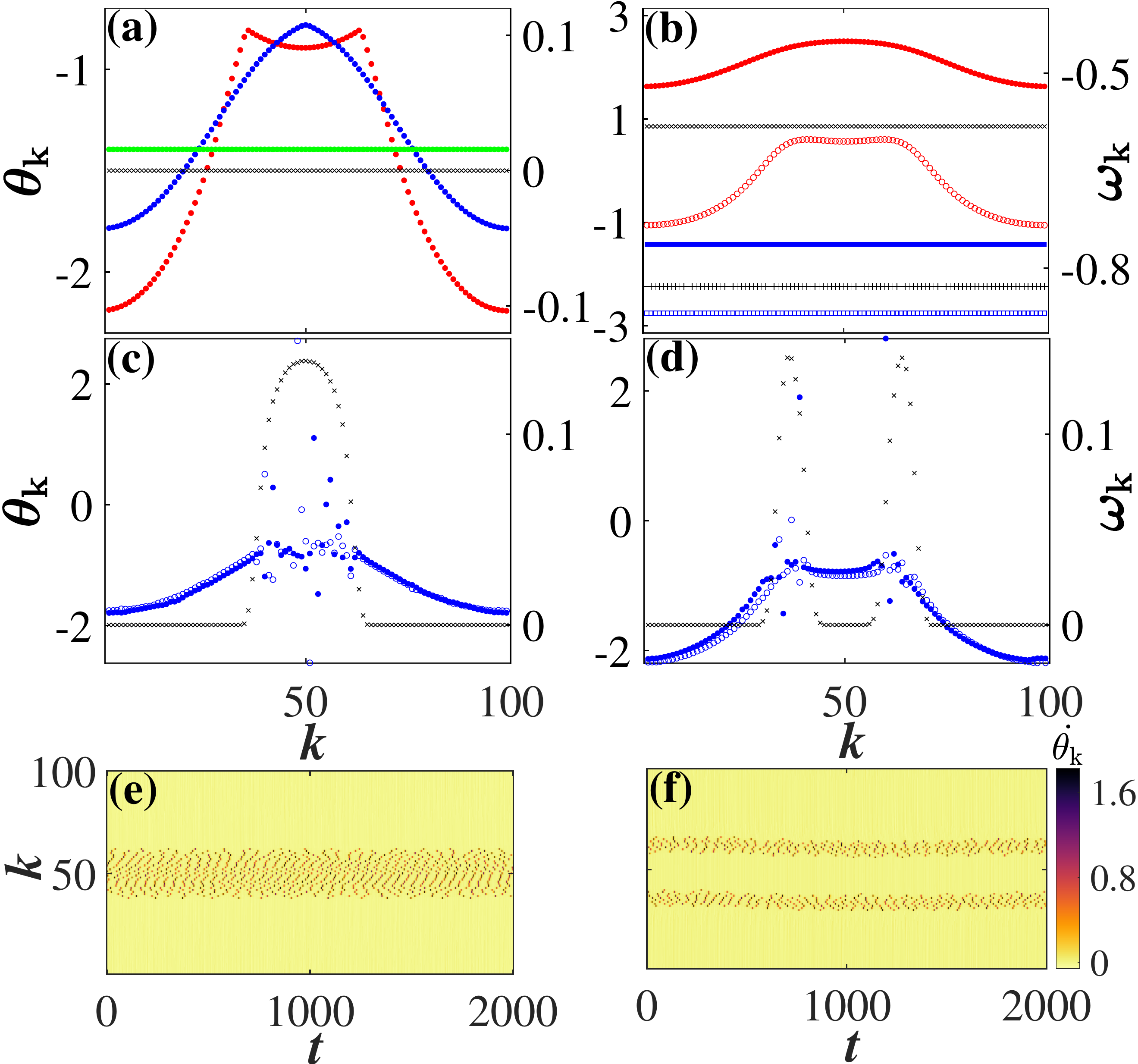}
\caption{Dynamical regimes of (\ref{Eq:Model}) with $N=100$, $P=35$, $\alpha=0.6$ and different choices of $K_1,\,K_2,\,a$. All the solutions were obtained from coherent spatially modulated initial conditions, and are plotted centered at $k=50$ using translational invariance. Snapshots of phases $\theta_k$ (colored symbols) and average frequencies $\omega_k$ (black) in (a)--(d).
(a)--completely coherent locked states for $a=1.2$: homogeneous (green) at $K_1=1.4,K_2=1.8$, spatially modulated (red, blue) at $K_1=1.4,K_2=1.983$ and $K_1=2,K_2=2.495$. (b)--time-periodic (unlocked) completely coherent states for  $a=0.5,K_2=1.8$: homogeneous (blue) at $K_1=3.4$, spatially modulated
(red) at $K_1=3.3$; different symbols of the same color indicate snapshots at different time moments. Bump states: (c)--single-headed for $K_1=1.4,K_2=2,a=1.2$, (d)--two-headed for $K_1=2,K_2=2.52,a=1.2$. Corresponding space-time plots of phase velocities $\dot{\theta_k}(t)$ in (e) and (f).
}\label{fig:states}
\end{figure}

In this Letter, we show that the solution to this outstanding problem can be found by applying the classical paradigm \cite{T52,GM72} of short-range activation and long-range inhibition to the synchrony in an array of coupled excitable/oscillatory units. We demonstrate that the coherence-incoherence patterns can emerge in a \emph{supercritical} scenario via a Turing instability of completely coherent states and a secondary homoclinic bifurcation, creating a \emph{single} incoherent oscillator, a state which can be seen as a \emph{weak chimera} in the sense of \cite{AB15}. Subsequent transitions via periodic, quasiperiodic and chaotic states with an increasing number of incoherent oscillators finally lead to a fully developed coherence-incoherence pattern with localized extensive chaos in the incoherent region. Remarkably, this scenario is essentially independent on the system size. 
We achieve this by introducing two modifications to the original phase oscillator system.
In addition to the nonlocal attractive coupling, we introduce a global repulsive coupling, and the uniformly rotating phase oscillators are replaced by so-called {\em active rotators}, which can be in an oscillatory or excitable regime. Coupled units of this type can be seen as a simplified version of neuronal oscillators similar to {\em theta neurons} \cite{L14,LBS13}, and under excitatory and/or inhibitory coupling they are known to display various types of localized or propagating spiking patterns. In particular, they can display localized states of activity, so-called
{\em bump states} \cite{OLC07,L11,L16,SA20,LO20}, which have also been extensively studied in continuum models for neuronal mean-field activity \cite{BGLM20,BAC19,C05,B12,LTGE02,LC01,WNCC14,MPR15,EARAM17}. With these additions, our system shows a variety of self-organized patterns,
see Fig.~\ref{fig:states}. Already completely coherent states, where all active rotators have identical average frequencies $\omega_k$, can be locked ($\omega_k=0$) or unlocked ($\omega_k\neq 0$), and in both cases they can be spatially uniform or spatially modulated, see Fig.~\ref{fig:states}(a) and (b), respectively. Incoherent regions, where average frequencies $\omega_k$ are gradually varying, can occur interspersed with locked coherent regions, as in bump states, see Fig.~\ref{fig:states}(c)-(d), or with unlocked coherent regions, as observed in chimera states.
The point that bumps and chimeras are related by a collective unlocking of the coherent region has previously been described in a system of active rotators even without global repulsion, having a classical chimera subjected to a global periodic forcing \cite{BSOP20}.

Along with a mechanism of emergence, a related puzzling aspect is that in their original form, chimeras cannot be observed in small systems. In \cite{WO11}, it has been shown that even for large system size, they are in fact chaotic transients collapsing to the completely coherent state after a lifetime that is exponentially increasing with the system size. There is of course no reason to believe that both these properties are necessarily true for any chimera-like phenomenon in systems other than Kuramoto's original phase oscillators. Indeed, Kuramoto's simple phase oscillator system allows for variations only in the phase lag parameter and the shape of the non-local coupling. Introducing of a more general coupling function has led to a discovery of weak chimera states \cite{AB15} which can also occur in rather small systems. However, they share only some of the properties of the classical chimeras, and it remained unclear to which extent the mechanisms of their emergence could serve as a general explanation of the original chimera phenomenon. In case of bump states, our results demonstrate that they can stably
exist in small systems without eventually collapsing to a completely coherent state.

We start with an array of $N$ oscillators where the dynamics of phases $\theta_j\in S^1$,
$j=1,\dots,N$ is given by
\begin{eqnarray}
\df{{\theta}_j}{t} &=&1 - a \cos\theta_j
- \fr{K_1}{2 P +1} \sum\limits_{k=j-P}^{j+P} \sin( \theta_j - \theta_k + \alpha ) \nonumber\\
&+& \fr{K_2}{N} \sum\limits_{k=1}^N \sin(\theta_j - \theta_k),\qquad j=1,\dots,N,
\label{Eq:Model}
\end{eqnarray}
where $K_1>0$ denotes the strength of the nonlocal attractive coupling and $K_2>0$ the global repulsive coupling.
For $a=0$, $K_2=0$, and an appropriate choice of the phase lag parameter $0<\alpha<\pi/2$ and the coupling range $1<P<N$,
this system is known to give rise to chimera states, see e.g. \cite{O13,WOS15,O18}.  Below $|a|=1$ the dynamics of individual
units changes from oscillatory to excitable.

\begin{figure}[ht]
\centering
\includegraphics[width=0.98\columnwidth]{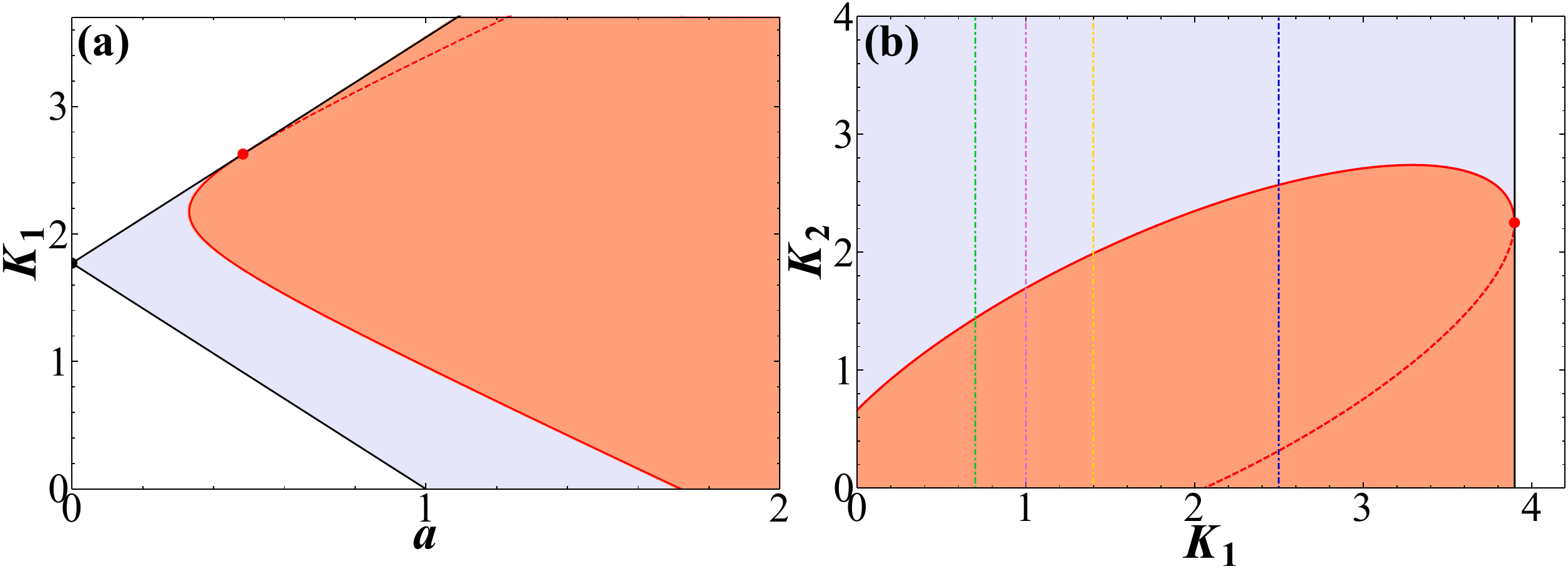}
\caption{Instabilities of the homogeneous locked state: fold \eqref{eq:TUBIF} with $\kappa=0$ (black), existence region (gray); Turing instability \eqref{eq:TUBIF}
for the mode with wave number $\kappa=1$ (red), stability region (orange); dashed parts of the curves lie on the unstable sheet.
(a)-- locking cone in the $(a, K_1)$ plane for fixed $K_2=1.4$; (b)-- locking region in the $(K_1,K_2)$ plane for fixed $a=1.2$.
Other parameters: $\alpha=0.6,N=100,P=35$. Vertical lines in (b) indicate choices of $K_1$ in Fig.~\ref{fig:reg}.
}\label{fig2}
\end{figure}

The system \eqref{Eq:Model} admits completely coherent homogeneous locked states
\begin{equation}
\theta_j(t)\equiv \theta^\pm = \pm\arccos\left[ \frac{1- K_1\sin\alpha}{a}\right],\quad 1\leq j\leq N, \label{eq:HE}
\end{equation}
which come in pairs within a locking cone
\begin{equation}
( K_1 \sin\alpha-1)^2<a^2, \label{eq:FOLD}
\end{equation}
with its tip located at $a=0$, $K_1=1/\sin\alpha$, cf. Fig. \ref{fig2}(a). Note that the locking cone does not depend on the global coupling $K_2$, since there is no phase lag in the corresponding coupling function. However, $K_2$ strongly affects the stability of the homogeneous locked states. Their Jacobian is a symmetric circulant matrix with real spectrum and discrete Fourier modes as eigenfunctions. The bifurcation condition for the mode with wave number $\kappa$ is given by
\begin{equation}
 a^2 = (K_1 (1 - R_\kappa)\cos\alpha + (\delta_{\kappa 0} -1)K_2 )^2 + (1 - K_1 \sin\alpha)^2, \label{eq:TUBIF}
\end{equation}
where
$$R_\kappa=\frac{1}{2P + 1}\sum_{m=-P}^P\cos(2\pi\kappa m/N)
$$
is the corresponding discrete Fourier component of the nonlocal coupling term. Note that inserting $\kappa=0$ into \eqref{eq:TUBIF} we recover the fold bifurcations outlining the locking cone \eqref{eq:FOLD}, while $\kappa=1,\dots N$ leads to a discrete Turing instability \cite{W12,CN20} with wave number $\kappa$. In Fig. \ref{fig2} we show the regions of existence and stability of the coherent uniform locked states. For $K_2=0$, the homogeneous locked state $\theta^-$ is stable within the whole locking region. Increasing $K_2$, the system undergoes a discrete Turing instability with the leading mode $\kappa=1$. If this bifurcation is supercritical, we obtain a stable spatially modulated completely coherent state (Turing pattern), see also \cite{KC18,CKR18}. We analyze now in detail four different destabilization scenarios of the homogeneous locked state induced by increasing the repulsive coupling $K_2$ along the vertical lines in Fig.~\ref{fig2}(b), which all finally lead to the onset of a bump state.

\paragraph{Sub- and Supercritical transitions to bump states}
\begin{figure}
\centering
\includegraphics[width=0.75\columnwidth]{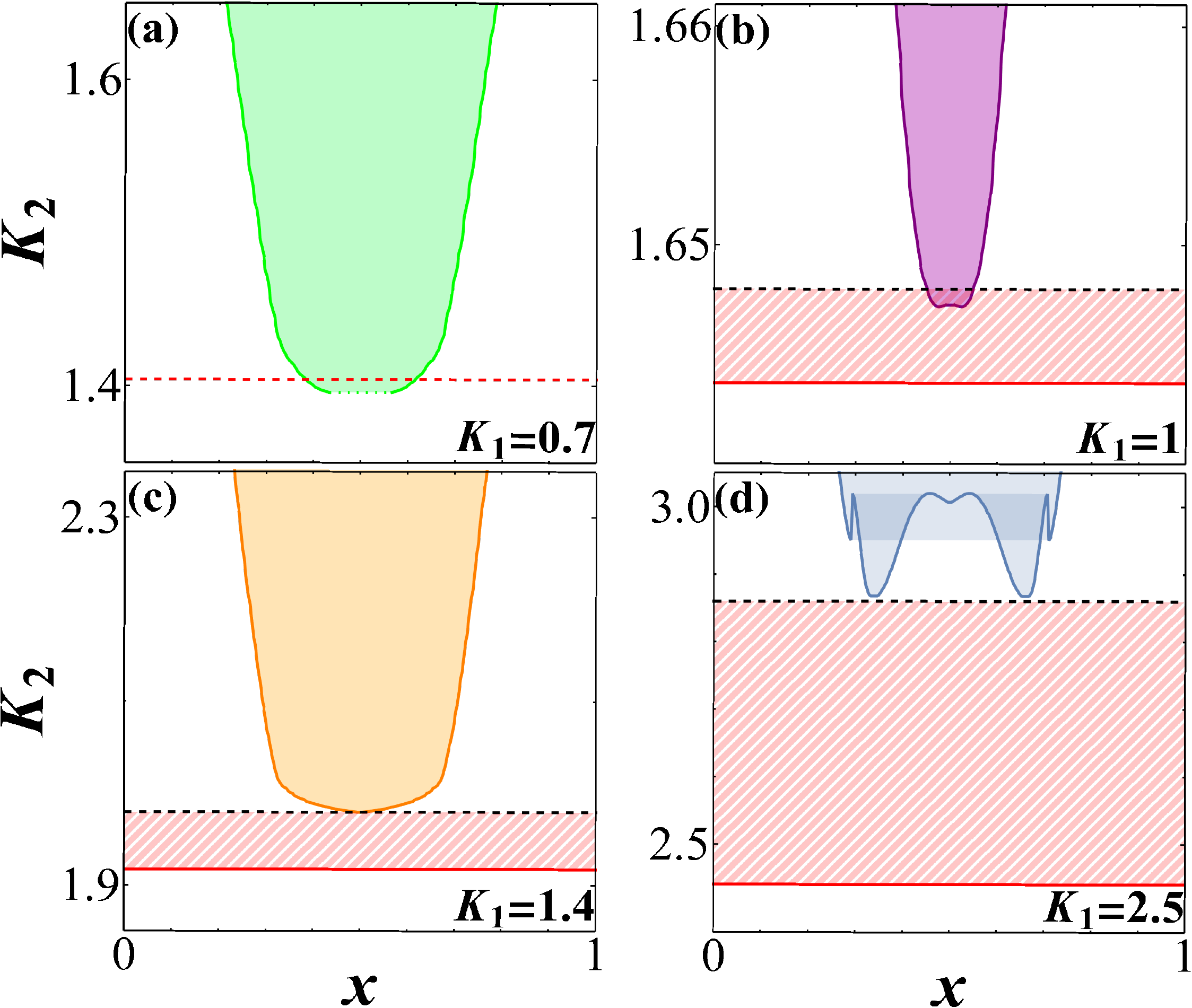}
\caption{Colored regions indicate increasing width of incoherent regions of bump states for varying $K_2$ and $K_1\in\{0.7,1,1.4,2.5\}$. Spatial parameter $x=k/N$ ($k$--oscillator index). Other parameters: $a=1.2$, $\alpha=0.6$. Turing instability of the homogeneous locked state (horizontal red line), where stable branches of modulated coherent locked states emerge (hatched region), ending at a saddle-node (black dashed line). If the saddle-node is a SNIC (cases (c) and (d)), there is a supercritical transition to bump states. In case (b) a saddle node of the modulated coherent state induces a subcritical transition to a fully developed bump. In case (a), the Turing instability is subcritical (red dashed line) and the solution jumps from the homogeneous coherent state to a bump.}\label{fig:reg}
\end{figure}
In Fig.~\ref{fig:reg} are illustrated different scenarios for the emergence of bump states showing how the incoherent region grows with increasing global repulsion $K_2$ for different choices of $K_1$. For larger values of $K_1$, see panels (b)-(d), the Turing bifurcation is supercritical and a branch of stable spatially modulated coherent states appears (hatched region). In (c) and (d), the stable branch of modulated coherent states extends to a SNIC (saddle-node on invariant circle) bifurcation. This instability represents the supercritical transition from a classical Turing pattern to a coherence-incoherence pattern. Remarkably, it is characterized by unlocking of single localized oscillators, independent on the system size $N$. Further increasing $K_2$ leads to the subsequent unlocking of neighboring oscillators and the coherence-incoherence pattern gradually attains temporal and spatial complexity. In Fig.~\ref{fig:reg}(d) is shown a scenario where the modulated coherent state develops two maxima, such that two incoherent regions emerge simultaneously. Increasing $K_2$ further, the two incoherent regions merge into a single one. During this process, the branch folds over and a region of coexistence of two different bump solutions appears. Fixing $K_1=1.4$, we observe a single monotonically growing incoherent region, as shown in Fig.~\ref{fig:reg}(c). This transition will be investigated in more detail below. For $K_1=1.0$, shown in panel (b), the saddle-node of the stable branch of modulated coherent states is a classical saddle-node bifurcation, which does not involve an invariant circle and is not localized to a single unlocking oscillator. Such a \emph{collective instability} can induce a subcritical transition to a coexisting fully developed bump state, with incoherent region of finite size. This transition shows a hysteretic behavior when the coupling parameter $K_2$ is reduced again, whereby the bump state disappears in a chaotic saddle before the size of the incoherent region completely vanishes. For small values of $K_1$ the Turing instability becomes subcritical, and there is a direct transition from the homogeneous coherent state to a fully developed bump, shown in panel (a), displaying the same hysteretic behavior as described above.
\begin{figure*}
\centering
\includegraphics[width=0.95\linewidth]{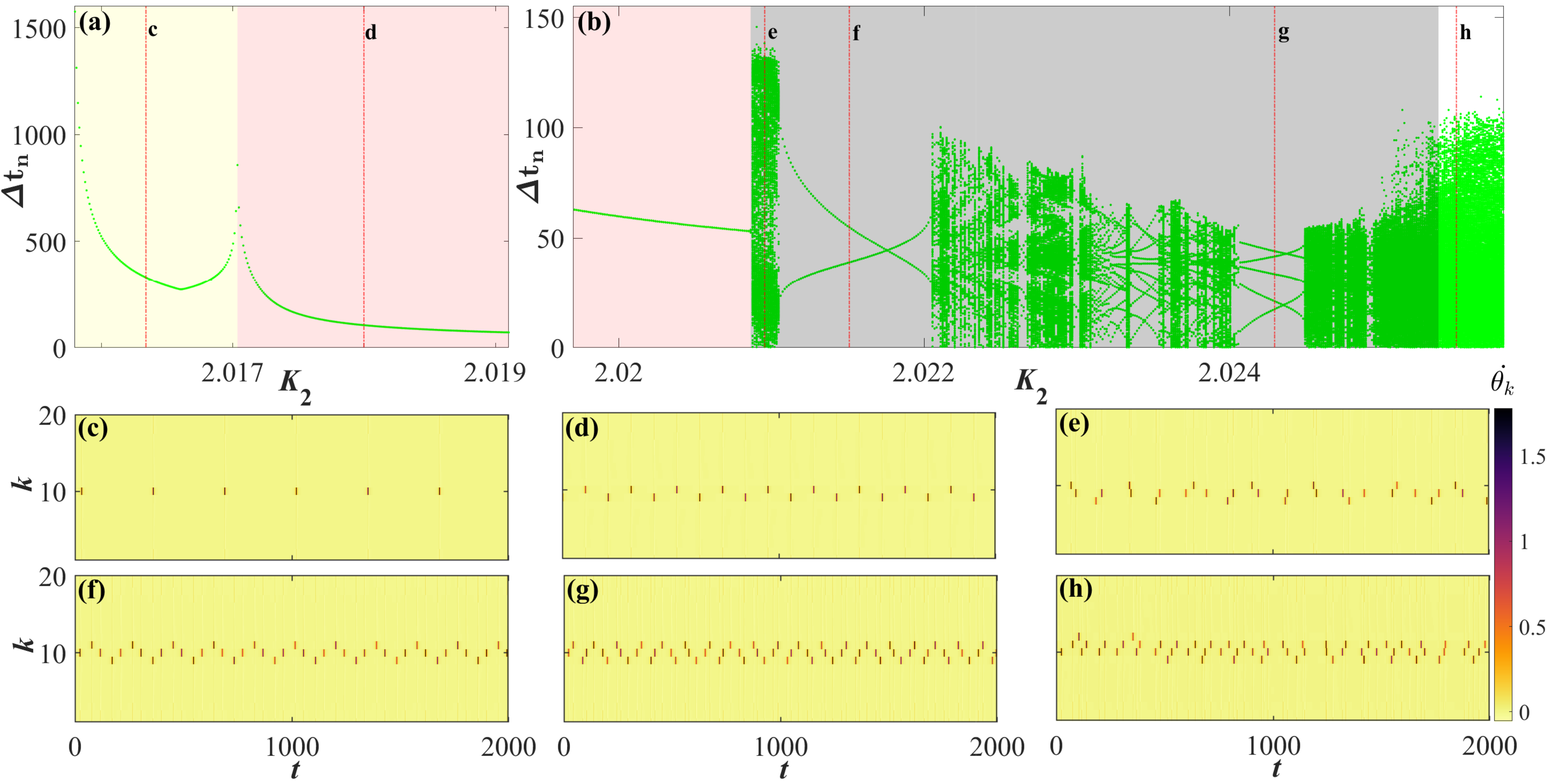}
\caption{(a)-(b) Bifurcation diagrams in $K_2$: time intervals $\Delta t_n$ between successive velocity peaks.
Space-time plots of phase velocities: periodic patterns in (c), (d), (f) and (g); chaotic patterns
without (e) and with drift instability (h) for $K_2$ values indicated by dash-dotted lines in (a) and (b). Other parameters
are $K_1=1.4,a=1.2,\alpha=0.6,N=20,P=7$.}\label{fig:bdiag}
\end{figure*}
\paragraph{Microscopic structure of the supercritical transition to bump states.} Directly after the SNIC bifurcation, when the number of incoherent oscillators in the bump states is small, one can observe an intricate scenario of increasing spatial and temporal complexity, which finally leads to
high dimensional extensive chaos. For large $N$, this transition is confined to a small parameter interval and one observes the almost immediate emergence of a small region of extensive chaos. In Figure \ref{fig:bdiag} we chose $N=20$ such that we can study in detail an example of such a transition. The resulting dynamics can be characterized by the spatio-temporal pattern of the single excitation events of the individual oscillators, which manifest themselves as localized peaks in the phase velocity. A selection of such patterns is given in panels (c)-(h) of Figure \ref{fig:bdiag}, while in panels (a) and (b) we show a full parameter scan with respect to $K_2$ where we sampled the return times $\Delta t_n$ between two consecutive peaks performed by any of the oscillators \cite{WOS15}. Starting from the simple periodic pattern with one incoherent oscillator that emerges from the SNIC of the modulated coherent state, we see multiple transitions between regular and chaotic states of increasing complexity. The transitions to chaos are mostly of intermittency type, but also torus breakup, illustrated in Fig.~\ref{fig:torus}, and period-doubling cascades can be observed. The shadings of different color in panels (a) and (b) indicates the increasing  number of incoherent oscillators. Note that for
$K_2\approx 2.0255$ the chaotic lateral motion of the incoherent region, which was described in \cite{WO11} for classical chimera states, sets in. Obviously, the specific shape of the transition scenario depends crucially on even small variations of the system parameters, in particular the number of oscillators $N$. However, a similar global scenario has been reported in \cite{WOS15}, where the classical chimera system of \cite{KB02} has been extended by a control term, such that also chimera states with a small number of incoherent oscillators became visible.
\begin{figure}[ht]
\centering
\includegraphics[width=0.85\columnwidth]{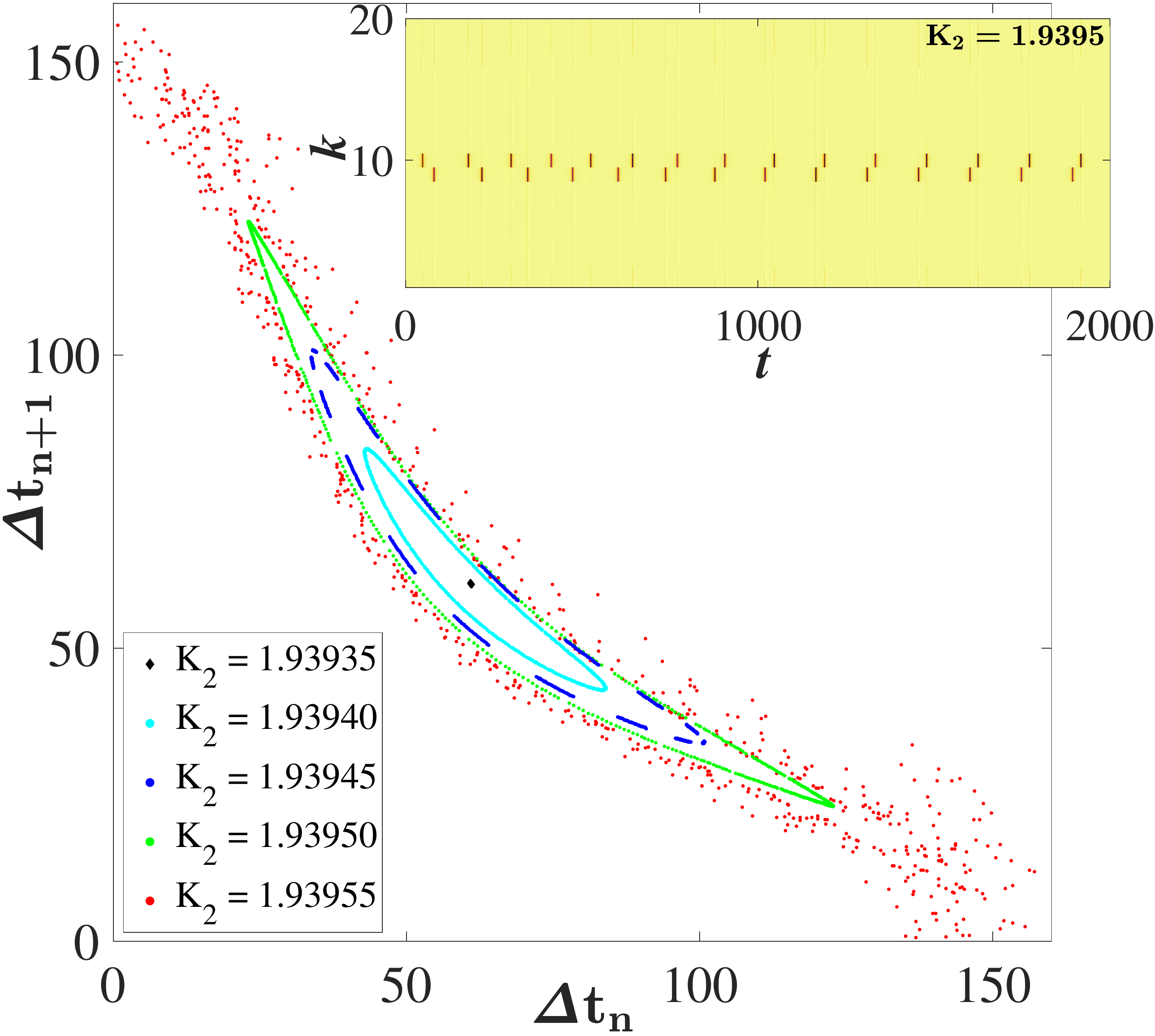}
\caption{Emergence of chaos under increasing $K_2$. Torus bifurcation from
periodic to quasiperiodic pattern at $K_2 \approx 1.9394$, onset of chaos via torus breakup at $K_2 \approx 1.93955$.
Inset: space-time plots of phase velocities for quasiperiodic pattern at $K_2=1.9395$.
Remaining parameters: $K_1=1.3,a=1.2,\alpha=0.6,N=20,P=7$.}\label{fig:torus}
\end{figure}

\paragraph{Outlook and discussion} Our system of excitable/oscillatory units with attractive and repulsive coupling, as given in (\ref{Eq:Model}), shows an extremely rich variety of dynamics. Here we have focused on the emergence of coherence-incoherence patterns and demonstrated how the classical paradigm of pattern formation by A. Turing \cite{T52} and Gierer \& Meinhardt \cite{GM72} in terms of local activation and long-range inhibition leads to the formation of coherence-incoherence patterns in a \emph{supercritical} transition scenario. In the spatially extended discrete medium of active rotators  the attractive and repulsive coupling with different spatial ranges does not activate/inhibit the local activity, as in neural field models, but acts on the local synchrony and in this way induces a pattern of qualitatively different behavior, rather than inducing quantitatively different levels of local activity as in the classical examples of pattern formation in neural field models. In this way, also the classical chimera states, which are related to the bump states discussed here by a simple collective unlocking of the coherent region as described in \cite{BSOP20}, are no longer an isolated phenomenon in the family of patterns, as stated in \cite{AS04}, but can be seen as a specific type of a Turing pattern, where a spatial modulation results in a self-localized unlocking that emerges gradually from a smooth coherent profile. We have further explained how this transition depends on the coupling strengths $K_{1,2}$. Moreover, we have shown that the coherence-incoherence patterns can be found for a large range of other parameters: in particular, the fine tuning of the phase lag $\alpha$ slightly below $\pi/2$ that was necessary to obtain chimera states in the classical setting \cite{O18,PA15} is no longer needed. Also, another puzzling aspect of chimera states in their original form could be resolved in our modified system. In \cite{WO11}, it was shown that chimeras cannot be observed in small systems, and that even for large system size, they behave as chaotic transients which collapse to the completely coherent state. With our extension of Kuramoto's simple phase oscillator system, coherence-incoherence patterns no longer need to coexist with the stable homogeneous state but, as we demonstrated, can be found as stable attractors even for small system size. In this way, sophisticated control schemes, which have been constructed for their observation \cite{SOW14,OOZS18}, become no longer necessary. Instead, our model is universal in the sense that the transition between the classical subcritical scenario and the new supercritical scenario
for the onset of coherence-incoherence patterns is achieved by the coupling parameters.


\begin{acknowledgments}
I.F. acknowledges funding from the Institute of Physics Belgrade through the grant by the Ministry of Education, Science and Technological Development of the Republic of Serbia. The work of O.O. was supported by the Deutsche Forschungsgemeinschaft under grant OM 99/2-1. The work of M.W. was supported by the Deutsche Forschungsgemeinschaft (DFG, German  Research  Foundation)--Projektnummer 163436311--SFB 910.
\end{acknowledgments}


\begin{thebibliography}{10}

\bibitem{KB02}{Y. Kuramoto, and D. Battogtokh, Nonlinear Phenom. Complex Syst. \textbf{5},380 (2002).}

\bibitem{AS04}{D. M. Abrams, and S. H. Strogatz, Phys. Rev. Lett. \textbf{93}, 174102 (2004).}

\bibitem{PJAS21}{F. Parastesh, S. Jafari, H. Azarnoush, Z. Shahriari, Z. Wang,
S. Boccaletti, and M. Perc, Phys. Rep. \textbf{898}, 1 (2021).}

\bibitem{PA15}{M. J. Panaggio and D. M. Abrams,
Nonlinearity \textbf{28}, R67 (2015).}

\bibitem{TRTSE18}{J. F. Totz, J. Rode, M. R. Tinsley, K. Showalter, and H. Engel,
Nature Phys. \textbf{14}, 282 (2018).}

\bibitem{HMRHOS12}{A. M. Hagerstrom, T. E. Murphy, R. Roy,
P. H{\"o}vel, I. Omelchenko, and E. Sch{\"o}ll,
Nature Phys. \textbf{8}, 658 (2012).}

\bibitem{LRCS19}{C. Lainscsek, N. Rungratsameetaweemana, S. S. Cash, and  T. J. Sejnowski,
Chaos \textbf{29}, 121106 (2019).}

\bibitem{ARMS16}
{R. G. Andrzejak, C. Rummel, F. Mormann, and K. Schindler,
Sci. Rep. \textbf{6}, 23000 (2016).}

\bibitem{OK19}
{O. E. Omel'chenko and E. Knobloch,
New J. Phys. \textbf{21}, 093034 (2019).}

\bibitem{AMSW08}{D. M. Abrams, R. Mirollo, S. H. Strogatz, and D. A. Wiley,
Phys. Rev. Lett. \textbf{101}, 084103 (2008).}

\bibitem{MBP16}{E. A. Martens, C. Bick, and M. J. Panaggio, Chaos \textbf{26}, 094819 (2016).}

\bibitem{MRJZ19}{M. Mikhaylenko, L. Ramlow, S. Jalan, A. Zakharova, Chaos \textbf{29}, 023122 (2019).}

\bibitem{Z20}{A. Zakharova, \emph{Chimera Patterns in Networks: Interplay Between
Dynamics, Structure, Noise, and Delay} (Springer International Publishing, 2020).}

\bibitem{OOHS13}{I. Omelchenko, O. E. Omel'chenko, P. H\"{o}vel, E. Sch\"{o}ll,
Phys. Rev. Lett. \textbf{110}, 224101 (2013).}

\bibitem{SZAS16}{N. Semenova, A. Zakharova, V. S. Anishchenko, E. Sch\"{o}ll,
Phys. Rev. Lett. \textbf{117}, 014102 (2016).}

\bibitem{HKK19}{K. H\"{o}hlein, F. P. Kemeth, and K. Krischer, Phys. Rev. E \textbf{100}, 022217 (2019).}

\bibitem{ZM21}
{Y. Zhang and A. E. Motter,
Phys. Rev. Lett. \textbf{126}, 094101 (2021).}

\bibitem{SK15}{L. Schmidt, and K. Krischer, Phys. Rev. Lett. \textbf{114}, 034101 (2015).}

\bibitem{T52}{A. M. Turing, Philos. Trans. R. Soc. B \textbf{237}, 37 (1952).}

\bibitem{GM72}{A. Gierer and H. Meinhardt,
Kybernetik {\bf 12}, 30 (1972).}

\bibitem{AB15}{P. Ashwin, and O. Burylko, Chaos \textbf{25}, 013106 (2015).}

\bibitem{L14}{C. R. Laing, Phys. Rev. E \textbf{90}, 010901(R) (2014).}

\bibitem{LBS13}{T. B. Luke, E. Barreto, and P. So, Neural Comput. \textbf{25}, 3207 (2013).}

\bibitem{OLC07}{M. Owen, C. Laing, and S. Coombes, New J. Phys. \textbf{9}, 378 (2007).}

\bibitem{L11}{C. R. Laing, Physica D \textbf{240}, 1960 (2011).}

\bibitem{L16}{C. R. Laing, Frontiers Comput. Neurosci. \textbf{10}, 53 (2016).}

\bibitem{LO20}{C. R. Laing, and O. E. Omel'chenko, Chaos \textbf{30}, 043117 (2020).}

\bibitem{SA20}
{H. Schmidt and D. Avitabile,
Chaos \textbf{30}, 033133 (2020).}

\bibitem{BGLM20}
{C. Bick, M. Goodfellow, C. R. Laing, and E. A. Martens,
J. Math. Neurosci. \textbf{10}, 9 (2020).}

\bibitem{BAC19}{\'{A}. Byrne, D. Avitabile, and S. Coombes, Phys. Rev. E \textbf{99}, 012313 (2019).}

\bibitem{C05}{S. Coombes, Biol. Cybern. \textbf{93}, 91 (2005).}

\bibitem{B12}{P. C. Bressloff, J. Phys. A Math. Theor. \textbf{45}, 033001 (2012).}

\bibitem{LTGE02}{C. R. Laing, W. C. Troy, B. Gutkin, and G. B. Ermentrout,
SIAM J. Appl. Math. \textbf{63}, 62 (2002).}

\bibitem{LC01}{C. R. Laing, and C. Chow, Neural Comput. \textbf{13}, 1473 (2001).}

\bibitem{WNCC14}{K. Wimmer, D. Q. Nykamp, C. Constantinidis, and A. Compte,
Nat. Neurosci. \textbf{17}, 431 (2014).}

\bibitem{MPR15}{E. Montbri\'{o}, D. Paz\'{o}, and A. Roxin,
Phys. Rev. X \textbf{5}, 021028 (2015).}

\bibitem{EARAM17}{J. M. Esnaola-Acebes, A. Roxin, D. Avitabile and E. Montbri\'o,
Phys. Rev. E \textbf{96}, 052407 (2017).}

\bibitem{BSOP20}{M. I. Bolotov, L. A. Smirnov, G. V. Osipov, and A. Pikovsky,
Phys. Rev. E \textbf{102}, 042218 (2020).}

\bibitem{WO11}{M. Wolfrum, and O. E. Omel'chenko, Phys. Rev. E \textbf{84}, 015201(R) (2011).}

\bibitem{O18}{O. E. Omel'chenko, Nonlinearity \textbf{31}, R121 (2018).}

\bibitem{O13}{O. E. Omel'chenko, Nonlinearity \textbf{26}, 2469 (2013).}

\bibitem{WOS15}{M. Wolfrum, O. E. Omel'chenko, and J. Sieber, Chaos \textbf{25}, 053113 (2015).}

\bibitem{W12}{M. Wolfrum, Physica D: Nonlinear Phenomena \textbf{241}, 1351 (2012).}

\bibitem{CN20}
{T. Carletti and H. Nakao
Phys. Rev. E \textbf{101}, 022203 (2020).}

\bibitem{CKR18}{C.-U. Choe, R.-S. Kim, and J.-S. Ri, Phys. Rev. E \textbf{98}, 012210 (2018).}

\bibitem{KC18}{R.-S. Kim, and C.-U. Choe, Phys. Rev. E \textbf{98}, 042207 (2018).}

\bibitem{SOW14}{J. Sieber, O. E. Omel'chenko, and M. Wolfrum,
Phys. Rev. Lett. \textbf{112}, 054102 (2014).}

\bibitem{OOZS18}{I. Omelchenko, O. E. Omel'chenko, A. Zakharova,
and E. Sch\"{o}ll, Phys. Rev. E \textbf{97}, 012216 (2018).}


\end{thebibliography}
\end{document}